\documentclass[preprint,12pt]{elsarticle}
\usepackage{graphicx}

\usepackage{amssymb}

\usepackage{amsthm}

\newtheorem{theorem}{Theorem}[section]
\newtheorem{lemma}[theorem]{Lemma}

\def\IC{{\mathbb C}}

\def\cV{{\mathcal V}}

\def\0{{\mathbf 0}}

\def\({\left(}
\def\){\right)}
\def\[{\left[}
\def\]{\right]}

\def\int{{\bf Int}}

\def\ra{{\rangle}}
\def\la{{\langle}}

\journal{Physics Letters A}

\begin{document}

\begin{frontmatter}


%
\title{Efficient Quantum Error Correction for Fully Correlated Noise}
%

\author{Chi-Kwong Li}
\ead{ckli@math.wm.edu}
\address{Department of Mathematics, College of William \& Mary,
Williamsburg, VA 23187-8795, USA. (Year 2011: Department of Mathematics, 
Hong Kong University of Science \& Technology, Hong Kong.)}

\author{Mikio Nakahara}
\ead{nakahara@math.kindai.ac.jp}
\address{Research Center for Quantum Computing,
Interdisciplinary Graduate School of Science and Engineering,
and Department of Physics,
Kinki University, 3-4-1 Kowakae, Higashi-Osaka, 577-8502, Japan.}

\author{Yiu-Tung Poon}
\ead{ytpoon@iastate.edu}
\address{Department of Mathematics, Iowa State University,
Ames, IA 50051, USA.}

\author{Nung-Sing Sze}
\ead{raymond.sze@inet.polyu.edu.hk}
\address{Department of Applied Mathematics, The Hong Kong Polytechnic 
University, Hung Hom, Hong Kong.}

\author{Hiroyuki Tomita}
\ead{tomita@alice.math.kindai.ac.jp}
\address{Research Center for Quantum Computing,
Interdisciplinary Graduate School of Science and Engineering,
Kinki University, 3-4-1 Kowakae, Higashi-Osaka, 577-8502, Japan.}

\begin{abstract}
We investigate an efficient quantum error correction of a
fully correlated noise. Suppose the noise is characterized by a quantum
channel whose error operators take fully correlated forms given by
$\sigma_x^{\otimes n}$, $\sigma_y^{\otimes n}$ and $\sigma_z^{\otimes n}$,
where $n>2$ is the number of qubits encoding the codeword. 
It is proved that (i) $n$ qubits codeword encodes $(n-1)$ data qubits when 
$n$ is odd and (ii) $n$ qubits codeword implements an error-free encoding,
which encode $(n-2)$ data qubits 
when $n$ is even. Quantum circuits implementing
these schemes are constructed.
\end{abstract}


\end{frontmatter}


\section{Introduction}

In quantum information processing, information is stored and processed 
with a quantum system. A quantum system is always in contact with its
surrounding environment, which leads to decoherence in the quantum system.
Decoherence must be suppressed for quantum information stored in qubits
to be intact. There are several proposals to fight against decoherence.
Quantum error correction, abriviated as QEC hereafter, is one of the
most promising candidate to suppress environmental noise, which leads to
decoherence \cite{gaitan}. By adding extra ancillary qubits, in analogy
with classical error correction, it is possible to encode a data qubit
to an $n$-qubit codeword
in such a way that an error which acted in the error quantum channel
is identified by measuring another set of ancillary qubits added for
error syndrome readout. Then the correct codeword is recovered from
a codeword suffering from a possible error by applying a recovery
operation, whose explicit form is determined by the error syndrome
readout.

In contrast with the conventional scheme outlined in the previous 
paragraph, there is a scheme in which neither syndrome readouts nor
syndrome readout ancilla qubits are required \cite{braunstein,laflamme,
tn,LNPST}. In particular, in \cite{tn,LNPST}, a general efficient 
scheme was proposed.
A data qubit is encoded with encoding ancilla qubits by the
same encoding circuit as the conventional one, after which a noisy
channel is applied on the codeword. Subsequently, the inverse of
the encoding circuit is applied on a codeword, which possibly suffers
from an error. The resulting state is a tensor product of the data
qubit state with a possible error and the ancilla qubit state.
It is possible to correct erroneous data qubit state by 
applying correction gates with the ancilla qubits as control
qubits and the data qubit as a target qubit.

This paper presents two examples of error correcting codes falling
in the second category. The noisy quantum
channel is assumed to be fully correlated \cite{chiri}, which means all
the qubits constituting the codeword are subject to the same
error operators. In most physical realizations of a quantum
computer, the system size is typically on the order of a
few micrometers or less, while the environmental noise, such as 
electromagnetic wave, has a wavelength on the order of a few
millimeters or centimeters.
Then it is natural to assume all the qubits in the register
suffer from the same error operator. 
To demonstrate the advantage of the
second category, we restrict ourselves within
the noise operators $X_n = \sigma_x^n, Y_n = \sigma_y^n, 
Z_n = \sigma_z^n$ in the following, where $n>2$ is the number of
constituent qubits in the codeword. We show that there exists an
$n$-qubit encoding which accommodates an $(n-1)$-qubit data state 
if $n$ is odd and an $(n-2)$-qubit date state if $n$ is even.
Although the channel is somewhat artificial as an error channel, 
we may apply our error correction
scheme in the following situation. Suppose Alice wants to send qubits to
Bob. Their qubit bases differ by unitary operations $X_n, Y_n$ or $Z_n$. 
Even when they do not know which basis the other party employs, 
the can correctly send qubits by adding one extra qubits (when $n$ is odd) or 
two extra qubits (when $n$ is even).

We state the theorems and prove them in the next section.
The last section is devoted to summary and discussions.

\section{Main Theorems}

In the following, $\sigma_i$ denotes the $i$th component of the Pauli
matrices and we take the basis vectors
$$
|0 \rangle = \left( \begin{array}{c}
1\\
0
\end{array} \right),\ \mathrm{and}\ |1 \rangle = \left( \begin{array}{c}
0\\
1
\end{array} \right)
$$
so that $\sigma_z$ is diagonalized. 
We introduce operators
$X_n=\otimes_{i=1}^n\sigma_x,\ Y_n=\otimes_{i=1}^n\sigma_y$ and 
$Z_n=\otimes_{i=1}^n\sigma_z$ acting on the 
$n$-qubit space $\IC^{2^n}=\otimes_{i=1}^n \IC^2$, where $n>2$
as mentioned before.

Let $A_1, A_2, A_3$ be $m\times m$ complex matrices, and let $k \in \{1, \dots, m-1\}$.
Denote by $\Lambda_k(A_1,A_2,A_3)$ the (joint)
rank-$k$ numerical range of $(A_1,A_2,A_3)$,
which is the collection of $(a_1, a_2, a_3) \in \IC^3$ such that $PA_jP = a_jP$ for some
$m\times m$ rank-$k$ orthogonal projection $P$ \cite{nr1,nr2,nr3}.
A quantum channel
of the form
\begin{equation}\label{eq:corrch}
\Phi(\rho) = p_0 \rho + p_1 X_n \rho X_n^{\dag}
+ p_2 Y_n \rho Y_n^{\dag}
+ p_3 Z_n \rho Z_n^{\dag}\ \hbox{ with } p_0, p_1, p_2, p_3 > 0, 
\ \sum_{i=0}^3 p_i = 1,
\end{equation}
has a $k$-dimensional quantum error correcting code (QECC) if and only if 
$\Lambda_k(X_n,Y_n,Z_n) \ne \emptyset$. To prove this statement, we need
to recall the Knill-Laflamme correctability condition, which asserts that
given a quantum channel $\Phi:M_n \to M_n$ with error operators $\{
F_i\}_{1 \leq i \leq r}$, 
$\cV$ is a QECC of $\Phi$ if and only if $P F_i^{\dagger} F_j P
= \mu_{ij} P$, where $P \in M_n$ is the projection operator with the
range space $\cV$ \cite{kl}. 
It should be clear that $\Lambda_k(\{F_i^{\dagger}
F_j\}_{1 \leq i,j \leq r}) \neq \emptyset$ if and only if there is a QECC
with dimension $k$. Now it follows from $X_n^2=Y_n^2=Z_n^2=I$ and the
relations 
$$
X_n Y_n = \pm Z_n,\ Y_n Z_n = \pm X_n,\ Z_n X_n = \pm Y_n
$$ 
when $n$ is even and 
$$
X_n Y_n = \pm i Z_n,\ Y_n Z_n = \pm i X_n,\ Z_n X_n = \pm i Y_n
$$
when $n$ is odd that the channel (\ref{eq:corrch}) has a 
$k$-dimensional QECC if and only if 
$$
\Lambda_k(\{F_i^{\dagger}F_j\}_{1 \leq i,j \leq r})
=\Lambda_k(X_n, Y_n, Z_n, I) \neq \emptyset. 
$$
By noting that $PIP = 1 \cdot P$ irrespective
of rank~$P$, we find $\Lambda_k(X_n, Y_n, Z_n) \neq \emptyset$
if and only if $\Lambda_k(X_n, Y_n, Z_n, I) \neq \emptyset$.

\begin{theorem} \label{thm1}   
Suppose $n>2$ is odd. Then  $\Lambda_{2^{n-1}}(X_n,Y_n,Z_n)\ne \emptyset$.
\end{theorem}

\noindent
\it Proof. \rm
Our proof is constructive.
For $j_1,\dots, j_n\in \{0,1\}$, denote
$|j_1,\dots, j_n\ra=\otimes_{i=1}^n|j_i\ra$. 
Let 
$$
\cV = {\mathrm{Span}}\left\{\, |j_1,\dots, j_n\ra:\mbox{the number of }i
\mbox{ with }j_i=1 \mbox{ is even} \right\}.
$$
Then
$\dim \cV=\sum_{r\mbox{ \scriptsize is even}} {n\choose r} =
\frac{1}{2}\((1+1)^n-(1-1)^n\)=
2^{n-1}$,
where ${n \choose r}$ is the number of $r$-combinations from $n$ elements.
Since 
$$
\sigma_x|0\ra=|1\ra, \ \sigma_x|1\ra=|0\ra, \ \sigma_y|0\ra=i|1\ra, 
\ \sigma_y|1\ra=-i|0\ra, \ \sigma_z|0\ra=|0\ra, \ \sigma_z|1\ra=-|1\ra,
$$ 
we have
$$X_n|v \rangle ,Y_n|v\rangle \in \cV^{\perp}\quad \hbox{and}\quad  Z_n|v
\rangle =|v \rangle  \quad \hbox{for all }|v \rangle \in\cV.$$
Let $P$ be the orthogonal projection onto $\cV$. Then the above observation
shows that $PX_nP=PY_nP=0$ and $PZ_nP=P$.
Therefore,  $(0,0,1) \in \Lambda_{2^{n-1}}(X_n,Y_n,Z_n)$, which shows
that $\Lambda_{2^{n-1}}(X_n,Y_n,Z_n) \ne \emptyset$ and hence
$\cV$ is shown to be a $2^{n-1}$-dimensional QECC.
\qed

\medskip
Now let us turn to the even $n$ case. We first state a lemma which
is necessary to prove the theorem.

\begin{lemma}\label{lem1}
Let $A \in M_N$ be a normal matrix. Then the rank-$k$ numerical range
of $A$ is the intersection of the convex hulls of any $N-k+1$ eigenvalues
of $A$.
\end{lemma}

The proof of the lemma is found in \cite{ls}.

\begin{theorem} \label{thm2}   
Suppose $n>2$ is even. Then  $\Lambda_{2^{n-2}}(X_n,Y_n,Z_n)\ne \emptyset$
but $\Lambda_{2^{n-1}}(X_n,Y_n,Z_n)= \emptyset$.
\end{theorem}

\noindent
\it Proof. \rm 
Let $n=2m$. By Theorem \ref{thm1},  
$\Lambda_{2^{n-2}}(X_{n-1},Y_{n-1},Z_{n-1})\ne \emptyset$. 
Consider 
$$
\cV' = {\mathrm{Span}}\left\{\, |0\rangle
|j_1,\dots, j_{n-1}\ra:\mbox{the number of }i
\mbox{ with }j_i=1 \mbox{ is even} \right\}.
$$
Observe that the projection $P$ onto $\cV'$ satisfies
$PX_nP=P Y_nP=0$ and $PZ_nP=P$ and hence $(0,0,1) \in 
\Lambda_{2^{n-2}}(X_n,Y_n,Z_n)$, which proves
$\Lambda_{2^{n-2}}(X_n,Y_n,Z_n)\ne \emptyset$. 

Since  $\{X_n,Y_n,Z_n\}$ is a 
commuting family, 
$X_n, Y_n$ and $Z_n$ can be diagonalized simultaneously. We may assume that
\begin{equation}\label{eq1}
X_n = I_{2^{n-1}}\oplus \left( -I_{2^{n-1}} \right) \quad\hbox{and}\quad 
Y_n = I_{2^{n-2}}\oplus \left( -I_{2^{n-2}} \right) \oplus I_{2^{n-2}} \oplus \left( -I_{2^{n-2}} \right).
\end{equation}
Since $\sigma_x\sigma_y=i\sigma_z$, we have
\begin{equation}\label{eq2} 
Z_n=(-1)^mX_nY_n=(-1)^m\(\, I_{2^{n-2}}\oplus \(-I_{2^{n-2}} \)\oplus \( -I_{2^{n-2}} \)\oplus I_{2^{n-2}}\, \).
 \end{equation}

Let us show that $\Lambda_{2^{n-1}}(X_n, Y_n)= \{(0,0)\}$.
We first note the identity $\Lambda_k(H, K) =
\Lambda_k(H+iK)$ for Hermitian $H, K$. Let us replace $H$ by $X_n$ and $K$
by $Y_n$ to obtain $\Lambda_k(X_n, Y_n) = \Lambda_k(X_n+iY_n)$. 
Since $X_n$ and $Y_n$ commute, $X_n+iY_n$ is normal and Lemma \ref{lem1}
is applicable. From Eqs.~(\ref{eq1}) and (\ref{eq2}), we find $X_n+iY_n$ has 
eigenvalues $1+i, 1-i, -1+i, -1-i$ and each eigenvalue is
$2^{n-2}$-fold degenerate. 
By taking $N=2^n$ and $k=2^{n-1}$ in Lemma \ref{lem1}, 
we find the rank-$2^{n-1}$ numerical range of $X_n+iY_n$ is the 
intersection of the convex hulls of any
$2^n-2^{n-1}+1=2^{n-1}+1$ eigenvalues. 
Since each eigenvalue has
multiplicity $2^{n-2}$, each convex hull involves at least three
eigenvalues. By inspecting four eigenvalues plotted in the complex
plane, we easily find the intersection of all the convex hulls
is a single point $(0,0)$, which proves
$\Lambda_{2^{n-1}}(X_n, Y_n)= \{(0,0)\}$. Similarly, we prove
$\Lambda_{2^{n-1}}(Y_n, Z_n)= \{(0,0)\}$. From these equalities
we obtain 
$$
\Lambda_{2^{n-1}}(X_n,Y_n,Z_n)\subseteq \{(0,0,0)\}.
$$
Suppose $\Lambda_{2^{n-1}}(X_n,Y_n,Z_n)\ne \emptyset$.
Let $P$ be a rank-$2^{n-1}$ projection such that
$PX_nP=PY_nP=PZ_nP=0$. Let $$P=\[\begin{array}{cc}P_{11}&P_{12}\\
 P_{12}^\dagger&P_{22}\end{array}\]$$
where each $P_{ij}$ has size $2^{n-1}\times 2^{n-1}$. 
From $P^2=P$ and $PX_nP=0$, we have four independent equations
$$
P_{11}^2+P_{12}P_{12}^\dagger=P_{11}, 
\ P_{11}^2-P_{12}P_{12}^\dagger=0, 
\ P_{22}^2+P_{12}^\dagger P_{12}=P_{22}, 
\ P_{22}^2-P_{12}^\dagger P_{12}=0. 
$$
Let $P_{12}=UDV^\dagger$ be the singular value decomposition of $P_{12}$,
where $D$ is a nonnegative diagonal matrix and 
$U, V \in {\mathrm{U}}(2^{n-1})$. Then
the above equations are solved as
$$
P_{11}=UDU^\dagger,\quad P_{22}=VDV^\dagger, \quad 2D^2=D.
$$
By collecting these results, we find the projection operator is decomposed as
$$
P=
\[\begin{array}{cc}U&0\\
0&V\end{array}\]
 \[\begin{array}{cc}D&D\\
 D&D\end{array}\]\[\begin{array}{cc}U^\dagger&0\\
0&V^\dagger\end{array}\]\,.
$$
Since rank~$P= 2^{n-1}$ and $P^2 = P$, 
it follows from $2D^2=D$ that  $D=\displaystyle \frac{1}{2}I_{2^{n-1}}$.
Let $$A=U^\dagger\(I_{2^{n-2}}\oplus (-I_{2^{n-2}})\)U
\quad\hbox{and}\quad
B=V^\dagger\(I_{2^{n-2}}\oplus (-I_{2^{n-2}})\)V\,.$$
Then both $A$ and $B$ are non-singular. On the other hand,
the assumption $ PY_nP =  PZ_nP= 0$ implies $A+B=A-B=0$ and
hence
$A=B=0$, which is a contradiction.
Therefore,  $\Lambda_{2^{n-1}}(X_n,Y_n,Z_n)= \emptyset$. \qed

\medskip
In the following,
we give an explicit construction of QECC for $\Phi$ in Eq.~(\ref{eq:corrch})
with odd $n$.
The technique is based on Theorem \ref{thm1} and the results in \cite{LNPST}.
Let $W$ be the $2^n \times 2^{n-1}$ matrix with columns in the set
$$
\{\, |j_1,\dots, j_n\ra:\mbox{the number of }i\mbox{ where }j_i=1 \mbox{ is even}\, \}.
$$

Define the $2^n \times 2^n$ matrix $R =
 \left[\begin{array}{cc} W & X_n W \end{array}\right]$.
In our QEC, an $(n-1)$-qubit state $\rho$
is encoded with one ancilla qubit $|0\ra$ as
$R(|0\ra\la0| \otimes \rho) R^\dag$. Then a noisy quantum channel $\Phi$
is applied on the encoded state and subsequently the recovery
operation $R^{\dag}$ is applied so that the decoded state 
automatically appears in the output with no syndrome measurements.
Our QEC is concisely summarized as
\begin{equation}\label{qecc}
R^\dag\, \Phi( R\, (|0\ra\la0| \otimes \rho)\,  R^\dag ) \, R = \rho_{a} \otimes \rho
\quad\hbox{for all}\quad \rho \in M_{2^{n-1}},
\end{equation}
where
$\rho_{a} = (p_0+p_3) |0 \ra\la 0| + (p_1+p_2) |1 \ra\la 1|$.

Choosing an encoding amounts to assigning each of $2^{n-2}$ column
vectors in $W$ a basis vector of the whole Hilbert space
without repetition. Therefore there are large degrees of freedom in
the choice of encoding. In the following examples, we have chosen
encoding whose quantum circuit can be implemented with the least number 
of CNOT gates. Since our decoding circuit is 
the inverse of the encoding circuit, it is also implemented with the least
number of CNOT gates.

When $n = 3$, the unitary operation $R$ can be chosen as
\begin{eqnarray*}
R
&=&|000\rangle\langle 000|+|011\rangle\langle 001|
+|110\rangle\langle 010|+|101\rangle\langle 011|\\
& &
+|111\rangle\langle 100|+|100\rangle\langle 101|
+|001\rangle\langle 110|+|010\rangle\langle 111|.
\end{eqnarray*}
When $n = 5$, $R$ can be chosen as
\begin{eqnarray*}
R
&=&|00000\rangle\langle 00000|+|00011\rangle\langle 00001|
+|00110\rangle\langle 00010|+|00101\rangle\langle 00011|\\
& &
+|01100\rangle\langle 00100|+|01111\rangle\langle 00101|
+|01010\rangle\langle 00110|+|01001\rangle\langle 00111|\\
& &
+|11000\rangle\langle 01000|+|11011\rangle\langle 01001|
+|11110\rangle\langle 01010|+|11101\rangle\langle 01011|\\
& &
+|10100\rangle\langle 01100|+|10111\rangle\langle 01101|
+|10010\rangle\langle 01110|+|10001\rangle\langle 01111|
\\
& &
+|11111\rangle\langle 10000|+|11100\rangle\langle 10001|
+|11001\rangle\langle 10010|+|11010\rangle\langle 10011|\\
&&
+|10011\rangle\langle 10100|+|10000\rangle\langle 10101|
+|10101\rangle\langle 10110|+|10110\rangle\langle 10111|\\
& &+|00111\rangle\langle 11000|+|00100\rangle\langle 11001|
+|00001\rangle\langle 11010|+|00010\rangle\langle 11011|\\
& &
+|01011\rangle\langle 11100|+|01000\rangle\langle 11101|
+|01101\rangle\langle 11110|+|01110\rangle\langle 11111|.
\end{eqnarray*}

\begin{figure}\label{nodd}
\begin{center}
\includegraphics[width=14cm]{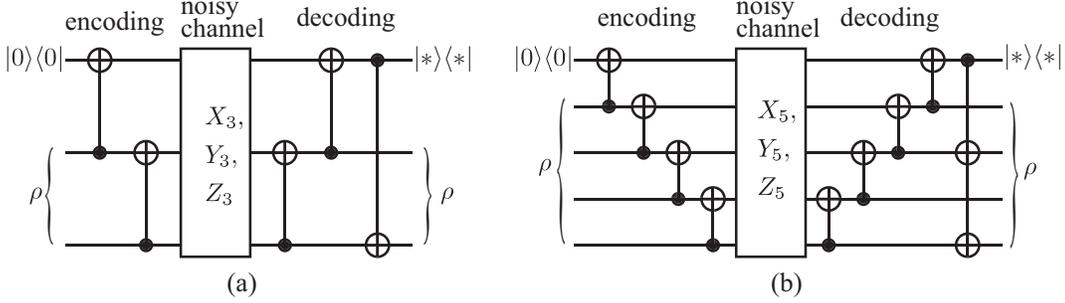}
\end{center}
\caption{Encoding and recovery circuits, which encodes and recovers
an arbitrary $(n-1)$-qubit state $\rho$ with a single ancilla qubit 
initially in the state
$|0 \rangle \langle 0|$. (a) is for $n=3$ while (b) is for $n=5$.
The quantum channel in the box represents
a quantum operation with fully correlated noise given in Eq.~(1).
The output ancilla state is $*=0\ (1)$ for error operators $I^{\otimes
3}$ and $Z_3$ ($X_3$ and $Y_3$) for $n=3$ and $*=0\ (1)$ for
$I^{\otimes 5}$ and $Z_5$ ($X_5$ and $Y_5$) for $n=5$.}
\end{figure}

Figure \ref{nodd}
shows quantum circuits of the matrix $R$ for $n=3$ and $n=5$. 
It follows from 
Eq.~(\ref{qecc}) that
the recovery circuit is the inverse of the encoding circuit.
It seems, at first sight, that
the implementations given in Fig.~1 contradict with 
Eq.~(\ref{qecc}) since the controlled NOT gate in the end of 
the recovery circuit
is missing in the encoding circuit.
Note, however, that the top qubit
is set to $|0\rangle$ initially and the controlled NOT 
gate is safely omitted without affecting encoding.

\begin{figure}\label{neven}
\begin{center}
\includegraphics[width=14cm]{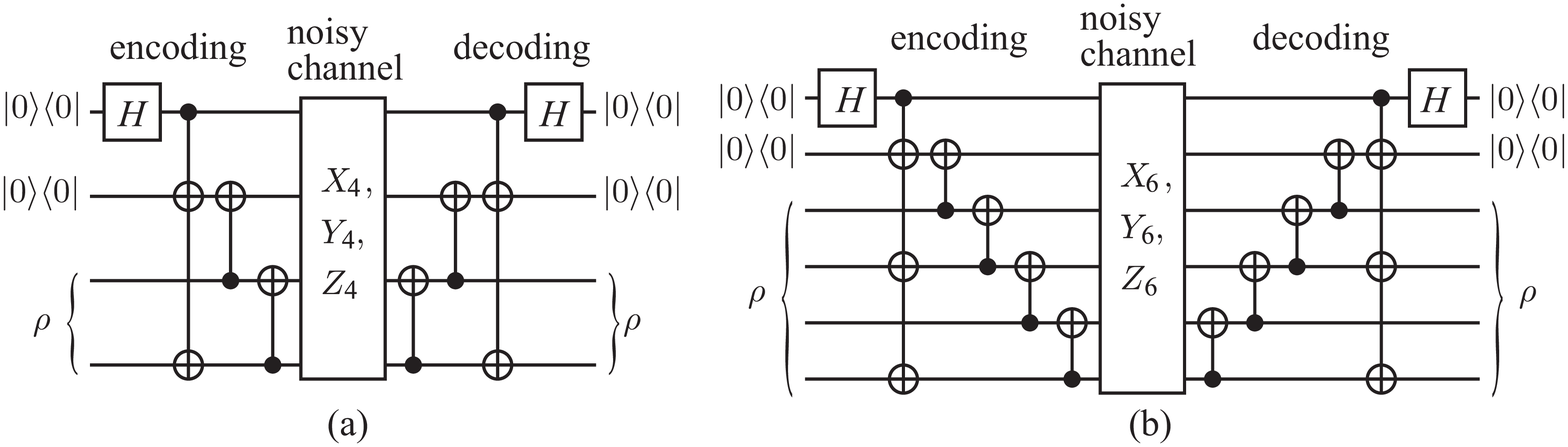}
\end{center}
\caption{Encoding and recovery circuits, which encodes and recovers
an arbitrary $(n-2)$-qubit state $\rho$ with two ancilla qubit 
initially in the state
$|00 \rangle \langle 00|$. (a) is for $n=4$ while (b) is for $n=6$.
The quantum channel in the box represents
a quantum operation with fully correlated noise given in Eq.~(1).
The output ancilla state is always $|00\rangle \langle 00|$, 
irrespective of error operators acted in the channel.}
\end{figure}

\medskip
We construct a decoherence-free encoding 
when $n$ is even as follows. The codeword in this case is immune to the noise
operators, which is an analogue of noiseless
subspace/subsystem introduced in \cite{ns,klv}.
Let
$$
|e\rangle=|j_1,\dots, j_n\rangle:
\mbox{the number of $i$ with $j_i=1$ is even}.
$$
Then evidently a vector
$$
\frac{1}{\sqrt{2}}~(|e\rangle~+~X_n|e\rangle)
$$
is separately invariant under the action of $X_n,~Y_n$ and $Z_n$.
There are
$$
\frac{1}{2}\sum_{r=\mbox{even}}\left(\begin{array}{c}n\\r\end{array}
\right)=2^{n-2}
$$
orthogonal vectors of such form, e.g. we have four vectors,
\begin{eqnarray}
&& \frac{1}{\sqrt{2}}(|0000\rangle+|1111\rangle),~
\frac{1}{\sqrt{2}}(|0011\rangle+|1100\rangle),\nonumber \\
&& \frac{1}{\sqrt{2}}(|0101\rangle+|1010\rangle),~
\frac{1}{\sqrt{2}}(|0110\rangle+|1001\rangle),
\label{nss}
\end{eqnarray}
for $n=4$.
Thus we find a decoherence-free encoding for $n-2=2$ qubits by projecting onto 
this invariant subspace spanned by these basis. It should be noted that
the projection operator $P$ to the subspace
$\cV_{\rm EF}$ spanned by the four vectors in Eq.~(\ref{nss}) satisfies
rank~$P=4$ and $PX_4P=PY_4P=PZ_4 P= P$, which shows 
$(1,1,1) \in \Lambda_4(X_4, Y_4, Z_4)$. 
It is easy to generalize this result to cases with arbitrary $n=2m >2$.
Figure \ref{neven} (a) and (b) depict quantum circuits for 
(a) $n=4$ and (b) $n=6$, respectively.

\section{Summary and Discussions}

We have shown that there is a quantum error correction which
suppresses
fully correlated errors of the form $\{\sigma_x^{\otimes n}, 
\sigma_y^{\otimes n}, \sigma_z^{\otimes n}\}$, in which
$n$ qubits are required to encode (i) $n-1$ data qubit states
when $n$ is odd and (ii) $n-2$ data qubit states when $n$ is even.
We have proved these statements by using operator theoretical
technique. 
Neither syndrome measurements nor ancilla qubits for syndrome 
measurement are required in our scheme, which makes physical
implementation of our scheme highly practical. 
Examples with $n=3$ and $n=5$ are analyzed in detail 
and explicit quantum circuits implementing our QEC with the least 
number of CNOT gate were obtained. 

Since the error operators are closed under matrix multiplication,
errors can be corrected even when they act on the codeword
many times.

A somewhat similar QEC has been reported in \cite{chiri}. 
They analyzed a partially correlated noise, where the
error operators acts on a fixed number of the codeword qubits 
simultaneously. They have shown that the quantum packing
bound was violated by taking advantage of degeneracy of
the codes. Justification of such a noise physically,
however, seems to be rather difficult. They have also 
shown that correlated noise acting on an arbitrary number $n$ of 
qubits can encode $k=n-2$ data qubits. In contrast, we have
analyzed a fully correlated noise, which shows the highest
degeneracy, and have shown 
that $k=n-1$ data qubits can be encoded
with an $n$-qubit codeword when $n$ is odd. 
Clearly, our QEC suppressing fully correlated errors is optimal as it is
clear that one cannot encode $n$ qubits as data qubits 
for odd $n$ and we have shown that one cannot encode $n-1$ qubits for
even $n$.

\section*{Acknowledgement}

CKL was supported by a USA NSF grant, a HK RGC grant,
the 2011 Fulbright Fellowship, and the 2011 Shanxi 100 Talent Program.
He is an honorary professor of University of Hong Kong,
Taiyuan University of Technology, and Shanghai University.
MN and HT were supported by ``Open Research Center'' Project
for Private Universities: matching fund subsidy from MEXT
(Ministry of Education, Culture, Sports, Science and Technology).
YTP was supported by a USA NSF grant.
NSS was supported by a HK RGC grant.



\bibliographystyle{model1a-num-names}







\end{document}